\documentclass[showpacs,aps,prx,floatfix,reprint,superscriptaddress,footinbib]{revtex4-2}

\usepackage{graphicx}
\usepackage{xfrac}
\usepackage{amssymb}
\usepackage{tikz}
\usepackage{amsfonts}
\usepackage{amsmath}
\usepackage[utf8]{inputenc}
\usepackage{mathtools}
\usepackage{xcolor,colortbl}
\usepackage{hyperref} \hypersetup{colorlinks=true,citecolor=blue,linkcolor=blue,urlcolor=blue}
\usepackage{bm}
\usepackage{array}
\usepackage{color}
\usepackage{relsize}
\usepackage[english]{babel}
\usepackage{amssymb,amsmath}
\usepackage{color}
\usepackage{pstricks}
\usepackage{graphicx}
\usepackage{color}
\usepackage{verbatim}
\usepackage{fancyvrb}
\usepackage{alltt}
\usepackage{longtable}
\usepackage{amsmath}

\definecolor{Gray}{gray}{0.9}
\definecolor{LightCyan}{rgb}{0.88,1,1}
\definecolor{green}{rgb}{0.2773,0.7422,0.3242}

\def\LDA{{\small  LDA}}
\def\GGA{{\small  GGA}}
\def\PBE{{\small  PBE}}

\def\AEL{{\small AEL}}

\def\DOS{{\small DOS}}

\def\AFLOWPOCC{{\small AFLOW-POCC}}
\def\AFLOWAPL{{\small AFLOW-APL}}
\def\POCC{{\small POCC}}

\def\QHPOCC{{\small QH-POCC}}
\def\KPPRA{{\small KPPRA}}
\def\QHA{{\small QHA}}
\def\EOS{{\small EOS}}
\def\LPS{{\small LPS}}
\def\AFLOW{{\small AFLOW}}

\def\VASP{{\small VASP}}

\def\PAW{{\small PAW}}
\def\ENMAX{{\small ENMAX}}

\def\sPOCC{{\substack{\scalebox{0.6}{POCC}}}}

\def\CTE{{\small CTE}}
\def\CTEs{{\small CTEs}}
\def\HNF{{\small HNF}}
\def\LTVC{{\small LTVC}}
\def\SQS{{\small SQS}}
\def\CPA{{\small CPA}}
\def\CE{{\small CE}}
\def\Veq{V_\textnormal{eq}}
\def\VeqT{\Veq(T)}

\def\kB{k_\textnormal{\tiny B}}
\def\CP{C_\textnormal{\tiny P}}
\def\CV{C_\textnormal{\tiny V}}
\def\Felec{F_\textnormal{elec}}
\def\Selec{S_\textnormal{elec}}
\def\Uelec{U_\textnormal{elec}}
\def\Fvib{F_\textnormal{vib}}
\def\fFD{f_\textnormal{FD}}
\def\EF{E_\textnormal{\tiny F}}
\def\gel{g_\textnormal{e}\left(\epsilon, V\right)}
\def\gph{g_\textnormal{ph}\left(\omega(V)\right)}
\def\Figure{Fig.}
\def\Figures{Figs.}

\def\BCT{{bct}}
\def\CUB{{cub}}
\def\ORC{{orc}}
\def\ORCC{{orcc}}
\def\MCLC{{mclc}}
\def\RHL{{rhl}}
\def\TET{{tet}}

\makeatletter \renewcommand\frontmatter@abstractwidth{\dimexpr\textwidth\relax} \makeatother

\def\memsduke{{Department of Mechanical Engineering and Materials Science, Duke University, Durham, North Carolina 27708, USA}}
\def\camd{{Center for Autonomous Materials Design, Duke University, Durham, North Carolina 27708, USA}}

\begin{document}

\title{\QHPOCC: taming tiling entropy in thermal expansion calculations of disordered materials}

\author{Marco Esters}
\thanks{These two authors contributed equally}
\affiliation{\memsduke} \affiliation{\camd}
\author{Andriy Smolyanyuk}
\thanks{These two authors contributed equally}
\affiliation{Institute of Solid State Physics, Technische Universit\"{a}t Wien, A–1040 Wien, Austria}
\affiliation{\camd}
\author{Corey Oses}
\affiliation{\memsduke} \affiliation{\camd}
\author{David Hicks}
\affiliation{\camd}
\author{Simon Divilov}
\affiliation{\memsduke} \affiliation{\camd}
\author{Hagen Eckert}
\affiliation{\memsduke} \affiliation{\camd}
\author{Xiomara Campilongo}
\affiliation{\camd}
\author{Cormac Toher}
\affiliation{Department of Materials Science and Engineering and Department of Chemistry and Biochemistry, University of Texas at Dallas, Richardson, Texas 75080, USA}
\affiliation{\camd}
\author{Stefano Curtarolo}
\email {email: stefano@duke.edu}
\affiliation{\camd} \affiliation{\memsduke}
\date{\today}

\begin{abstract}
\noindent
Disordered materials are attracting considerable attention because of their enhanced properties
compared to their ordered analogs, making them particularly suitable for high-temperature
applications. The feasibility of incorporating these materials into new devices depends on a variety
of thermophysical properties. Among them, thermal expansion is critical to device stability, especially in multi-component
systems. Its calculation, however, is quite challenging for materials
with substitutional disorder, hindering computational screenings.
In this work, we introduce \QHPOCC\ to leverage the local
tile-expansion of disorder. This method provides an effective partial
partition function to calculate thermomechanical properties of substitutionally disordered compounds in the quasi-harmonic
approximation.
Two systems, AuCu$_3$ and CdMg$_3$, \textcolor{black}{the latter a candidate for long-period
superstructures at low temperature}, are used to validate the methodology by
comparing the calculated values of the coefficient of thermal expansion and isobaric heat capacity
with experiment, demonstrating that \QHPOCC\ is a promising approach to study thermomechanical
properties of disordered systems.
\end{abstract}

\maketitle

\section*{Introduction}
\label{intro}
\noindent
Multi-component alloys~(MCAs) and ceramics~(MCCs) are novel classes of materials offering enhanced
properties such as
ultra-high hardness~\cite{curtarolo:art140, Wang_HECarbides_AdvTheorySimul_2020},
low thermal conductivity~\cite{Braun_ESO_AdvMat_2018},
and good corrosion and wear resistance~\cite{Tsai_HEAReview_MRL_2014,
Ye_hea_high_melting_point, HE_ceramics_review}.
The combination of these properties makes them suitable for applications in
thermal barrier coatings~\cite{Zhou_HEZirconate_JEurCeramSoc_2020},
thermoelectrics~\cite{Jiang_HEChalcogenides_Science_2021},
ultra-high temperature structural applications~\cite{Wuchina_UHTCs_borides_carbides_nitrides_2007},
and diffusion barriers for microelectronics~\cite{Tsai_Nitride_ApplPhysLett_2008}.
In these types of applications, a material is often part of a multi-component system. It is thus not
only critical for the material to have optimized properties, but it must
also be compatible with the remaining parts of the device at operating
conditions \cite{curtarolo:art80}.
For example, combining components with incompatible coefficiencts of thermal expansion~(\CTEs)
would compromise the integrity of the product at higher temperatures.

While measuring \CTEs\ has become an increasingly integral part of experimental research of
MCAs and MCCs~\cite{Laplanche_George_JAC_2018, Zhu_HENiobates_JEurCeramSoc_2021,Zhu_HEZirconate_JEurCeramSoc_2021, Chen_HENiobates_ApplPhysLett_2021},
the substitutional disorder presents a formidable challenge for computational investigations
of these materials, resulting in only few computed \CTE\ values with scarce experimental
validation~\cite{Ma_ActaMater_2015,Huang_AplPhysLet_2017}.
To date, calculations of thermal expansion in systems with
substitutional disorder have employed the
Coherent Potential Approximation~(\CPA)~\cite{Soven_PhysRev_1967, Gyorffy_PRB_1972},
Special Quasirandom Structures~(\SQS)~\cite{sqs}, or
the cluster expansion~(\CE) method~\cite{sanchez_ducastelle_gratias:psca_1984_ce,deFontaine_ssp_1994}.
The \CPA\ allows for modeling alloys of any stoichiometry using relatively small unit cells.
However, it cannot reliably calculate atomic forces~\cite{Vitos_Springer_2007}
and is thus unable to employ conventional lattice dynamics methods.
\SQS\ models disordered systems at a fixed composition by finding the most
random~(infinite temperature) representation of the structure. As a consequence, it neglects
finite-temperature short-range ordering effects.
\CE\ calculates thermodynamic properties through a series expansion of the free energy using a set
of ordered structures. While the obtained free energy is exact, the combinatorial explosion of the
number of cluster expansion parameters makes the study of complex MCAs and MCCs impractical.

The Partial OCCupation~(\POCC) method was developed to include finite-temperature effects while
alleviating the computational cost of \CE~\cite{curtarolo:art110},
thus allowing practical studies of multi-component systems.
\POCC\ models the disordered material
through a series of small {\it tiles} --- ordered representative supercells with the same stoichiometry as the
parent structure ---  which are also capable of calculating vibrational properties~\cite{curtarolo:art180}.
This particular expansion allows \POCC\ to factorize the
{\it tiling entropy} associated with the latent heat of a first order phase
transition (from chemical order and microstrucural disorder to
chemical disorder and no microstrutures), and
to generate a partial partition function out of the global one. The
approach, taken at finite temperatures with the quasi-harmonic
approximation~(\QHA), enables the calculation of thermomechanical
properties.
The extended method, called \QHPOCC, is tested on AuCu$_3$ and CdMg$_3$
for the ordered and disordered fields.
Both thermal expansion and isobaric heat capacity are in good agreement with experiments, demonstrating the potential of
\QHPOCC\ to accurately model thermomechanical properties in the \QHA\ at a reasonable computational
cost.
\textcolor{black}{We chose the latter compound for its capabilities in forming long-period superstructures at low temperature
\cite{Massalski,monster,curtarolo:art54}, so that the effect of the diverging characteristic lengths of the
ground-state can be qualitatively compared against the finite size of the \POCC\ tiles expansion.
}

\section*{Computational methods}

{\bf Thermal Expansion of Disordered Systems.}
The volumetric \CTE\ $\beta$ is defined as the logarithmic derivative of the
temperature-dependent equilibrium volume:
\begin{equation}
  \beta = \frac{d\log(\VeqT)}{dT} = \frac{1}{\VeqT} \frac{d \VeqT}{d T},
  \label{eq:cte}
\end{equation}
where $\VeqT$ is the equilibrium volume at temperature $T$, i.e., the volume $V$ for which the
temperature- and volume-dependent free energy $F$ is at a minimum:
\begin{equation}
  \frac{\partial F(V,T)}{\partial V} = 0.
  \label{eq:fvt}
\end{equation}
$F(V, T)$ consists of the following contributions when neglecting anharmonic effects:
\begin{equation}
  \label{eq:F_derivative_structure}
  F (V, T) = E_0(V) + \Felec(V, T) + \Fvib(V, T).
\end{equation}
$E_0$ is the energy of the structure at 0\,K. $\Felec$ is the electronic contribution to the free
energy and can be calculated as:
\begin{align}
\Felec(V, T) &= \Uelec - T\Selec,\label{eq:Felec}\\
\Uelec(V, T) &= \int_0^\infty \fFD\gel\epsilon d\epsilon - \int_0^{\EF} \gel\epsilon d\epsilon,\label{eq:Uelec}\\
\Selec(T) &= -\kB\int_0^\infty \gel s_T(\epsilon) d\epsilon.\label{eq:Selec}
\end{align}
Here, $\gel$ is the electronic density of states (\DOS), $\fFD = \fFD\left(\epsilon, T\right)$ is
the Fermi-Dirac distribution, $\EF$ is the Fermi energy, and
$s_T(\epsilon) = \fFD\log(\fFD) + (1 - \fFD)\log(1 - \fFD)$ is the electronic entropy of $\epsilon$.

The last contribution, $\Fvib$, is the vibrational free energy. It can be calculated via the phonon
\DOS, $\gph$, as:
\begin{equation}
  \Fvib(V, T) = \kB T \int_0^\infty\log\left(2\sinh\frac{\hbar\omega(V)}{2\kB T}\right)\gph d\omega,
  \label{eq:Fvib}
\end{equation}
where $\omega$ is the phonon frequency and $\hbar$ is the reduced Planck constant.

In the \QHA, $\VeqT$ is obtained by calculating $F(V,T)$ over a set of volumes and fitting the
volume-dependent free energy to an equation of state~(\EOS) for each temperature.
Calculating the volume-dependency of $F$ is thus the central problem for the calculation of
thermomechanical properties at finite temperatures.

Computational modeling of MCAs and MCCs --- and therefore accurate
calculation of free energies ---
is challenging due to the substitutional disorder in these compounds.
\AFLOWPOCC\ describes
disordered materials through an ensemble of small Hermite-normal-form~{\color{blue}(\HNF)}
ordered representatives~{\color{blue}\cite{gus_enum,Santoro_HNF1,Santoro_HNF2}},
also called {\it tiles}.
This method reproduces the correct composition, symmetry, and properties
that depend on the radial distribution function of the system but needs
to be integrated over all possible {\it tiling configurations} if the
global partition function $Z_{\rm global}$ and free energy
$F_{\rm global}$ are needed. For an $NVT$ ensemble, $Z_{\rm global}$ can
be calculated as:
\begin{equation}
  Z_{\rm global}\left(V, T\right) \equiv \!\!\!\!\!\!\!\!\!\!\!  \sum_{\substack {i -{\rm tiles}\\
      {\rm tiling}\left[\left\{V_i\right\}\right]=V}}\!\!\! \! \!\! g_i \exp\left(-\frac{F_i\left(V_i, T\right)}{\kB T}\right),
  \label{eq:ZNVT_global}
\end{equation}
where $\kB$ is the Boltzmann constant and  $V_i$ is the volume of the $i^\textnormal{th}$ tile.
The sum is constrained by the organization of the tiles covering
the whole normalized volume $\sum \left\{V_i\right\}/N=V$ with the global number of tiles $N$.
$Z_{\rm global}\left(V, T\right)$ can be used to obtain $F_{\rm global}$ through:
\begin{equation}
  F_{\rm global}(V,T) \equiv -\kB T \log Z_{\rm global}\left(V_i, T\right).
  \label{eq:F_global}
\end{equation}

\begin{figure}
  \centering
  \includegraphics[width=\linewidth]{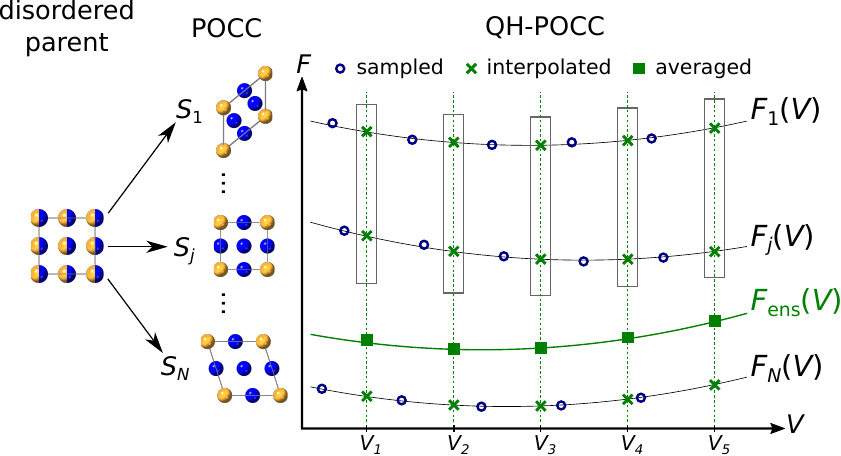}
  \caption{
    {\bf Illustration of the \QHPOCC\ workflow.}
    The free energies of the \AFLOWPOCC\ tiles $\{S_i\}$
    are sampled for various volumes near the equilibrium.
    The free energies are interpolated and then ensemble-averaged
    to resolve the free energy of the disordered system,
    from which thermomechanical properties can be derived.
  }
  \label{fig:Ensemble_Illustration}
\end{figure}

The calculation of $Z_{\rm global}\left(V, T\right)$ is highly non-trivial.
The number of possible configurations $\Omega_{\rm tiling}$ --- having the correct
volume where all the pieces fit together --- increases with the cut off
of the maximum \POCC\ tile-size.
Thermodynamically, this number also leads to the tiling entropy
$S_{\rm tiling} \equiv \kB \log \Omega_{\rm tiling}$.
The properties describing the ``organization'',
$\left\{\Omega_{\rm tiling}, S_{\rm tiling} \right\}$,
can be calculated combinatorially or through Monte Carlo modeling like in percolation or
self-avoiding random-walks theories, but incurs great computational cost.
Here, these quantities can be neglected because of the following considerations:\\
{\bf i.}
$\Omega_{\rm tiling}$ varies slowly unless near a first-order phase transition.
The number of possible configurations depends on the identity and distribution of the
individual tiles. Therefore, $\Omega_{\rm tiling}$ is nearly constant
below and above the transition temperature, albeit not the same in general.
At the phase transition, on the other hand, the Boltzmann population vector suddenly
rotates in the probability hyperspace~(see Fig. 2e in Ref. \cite{aflowLTVC}), leading to
completely different tiling-distribution identities and thus a different $\Omega_{\rm tiling}$.
Similar considerations hold for the tiling entropy:
$S_{\rm tiling}(T)$ is a slowly varying function of $T$ unless a phase
transition is crossed.\\
{\bf ii.}
Derivatives $\partial [\cdot ]/ \partial T$ operating on functionals containing
$S_{\rm tiling}(T)$ may ignore tiling contributions.
As long as $\partial [\cdot]/ \partial T$ is not operated
at the phase transition (or nearby, as precursors start appearing~\cite{aflowLTVC}),
$S_{\rm tiling}(T)$ can be neglected with great computational savings.
Temperature-derivative properties, such as specific heat and thermal expansion,
can be estimated below and above the transition temperature without direct
knowledge of the organization of the tiles.\\
{\bf iii.}
Once the tiling organization is settled, phase stability requires
minimization of free energies. This, in turn, requires minimization with respect
the total tiling organization along the
aforementioned volume constraints,
$\sum \left\{V_i\right\}/N=V$
(Lie derivative).
In general, $\min \left[ \sum \left\{\cdot \right\} \right] \neq \sum
\min \left[ \left\{\cdot \right\}\right]$ --- yet, the \POCC\ model allows to solve
this conundrum: in canonical \POCC-tiles, both species concentration and superlattice are
conserved. Only atoms are swapped generating symmetrically inequivalent
decorations~\cite{curtarolo:art110}.
As such, the volumes $V_i$ and minimum energy $E_i$~(or $H_i$ in an applied stress field)
for each $i$-tile will be quite close, $V_i\sim V$, so that the global \EOS\ can be summarized as:
\begin{eqnarray}
  \min&&\left[{\rm EOS}\{V_i\}\right] \equiv
         \min \left[                                \sum_i {{\rm EOS}_i}(V_i) \right] \approx \nonumber \\
      &&\approx \sum_i \min \left[ {\rm EOS}_i (V_i) \right] \approx \sum_i {\rm EOS}_i (V).  \label{eq:EOS_approx}
\end{eqnarray}
Equation (\ref{eq:EOS_approx}) coupled with Equations
(\ref{eq:ZNVT_global}) and (\ref{eq:F_global}) allow the definition of
a partial partition function $Z_{\rm \sPOCC}(V,T)$ and free energy $F_{\rm \sPOCC}(V,T)$:
\begin{eqnarray}
  Z_{\rm \POCC}\left(V, T\right)  &\equiv& \sum_{i-{\rm tiles}}g_i \exp\left(-\frac{F_i\left(V,T\right)}{\kB T}\right), \label{eq:ZNVT_partial} \\
  F_{\rm \POCC}(V,T)&\equiv& -\kB T \log Z_{\rm \sPOCC}\left(V, T\right).
  \label{eq:F_partial}
\end{eqnarray}
Even though $Z_{\rm \sPOCC}$ and $F_{\rm \sPOCC}$ neglect tiling entropy,
they are capable of reproducing $T$-derivative properties outside regions of
phase transition, without the cumbersome calculation of the global partition function.
 Equation (\ref{eq:F_partial}) can then be used to construct the \EOS\ of
of the disordered material, leading to the global coefficient of thermal expansion via
Equation (\ref{eq:cte}). This scheme is illustrated in \Figure~\ref{fig:Ensemble_Illustration}.

From the \EOS-fitted curve, other thermomechanical properties can be derived. The isobaric ($\CP$)
and isochoric ($\CV$) heat capacities can be calculated as:
\begin{align}
  \CP &= -T\left(\frac{\partial^2 F_{\rm \sPOCC}(V,T)}{\partial T^2}\right)_P = -T\frac{\partial^2 F_{\rm \sPOCC}(\Veq,
  T)}{\partial T^2},\label{eq:Cp}\\
  \CV &= \CP - \Veq B \beta^2 T,\label{eq:Cv}
\end{align}
where $B$ is the bulk modulus:
\begin{equation}
  B = V \frac{\partial^2 F_{\rm \sPOCC}(V, T)}{\partial V^2}.
  \label{eq:B}
\end{equation}
These quantities can be used to determine the effective Gr\"{u}neisen parameter $\bar\gamma$:
\begin{equation}
  \bar\gamma = \frac{\Veq B \beta}{\CV}.
  \label{eq:Grueneisen}
\end{equation}

\begin{figure*}
  \includegraphics{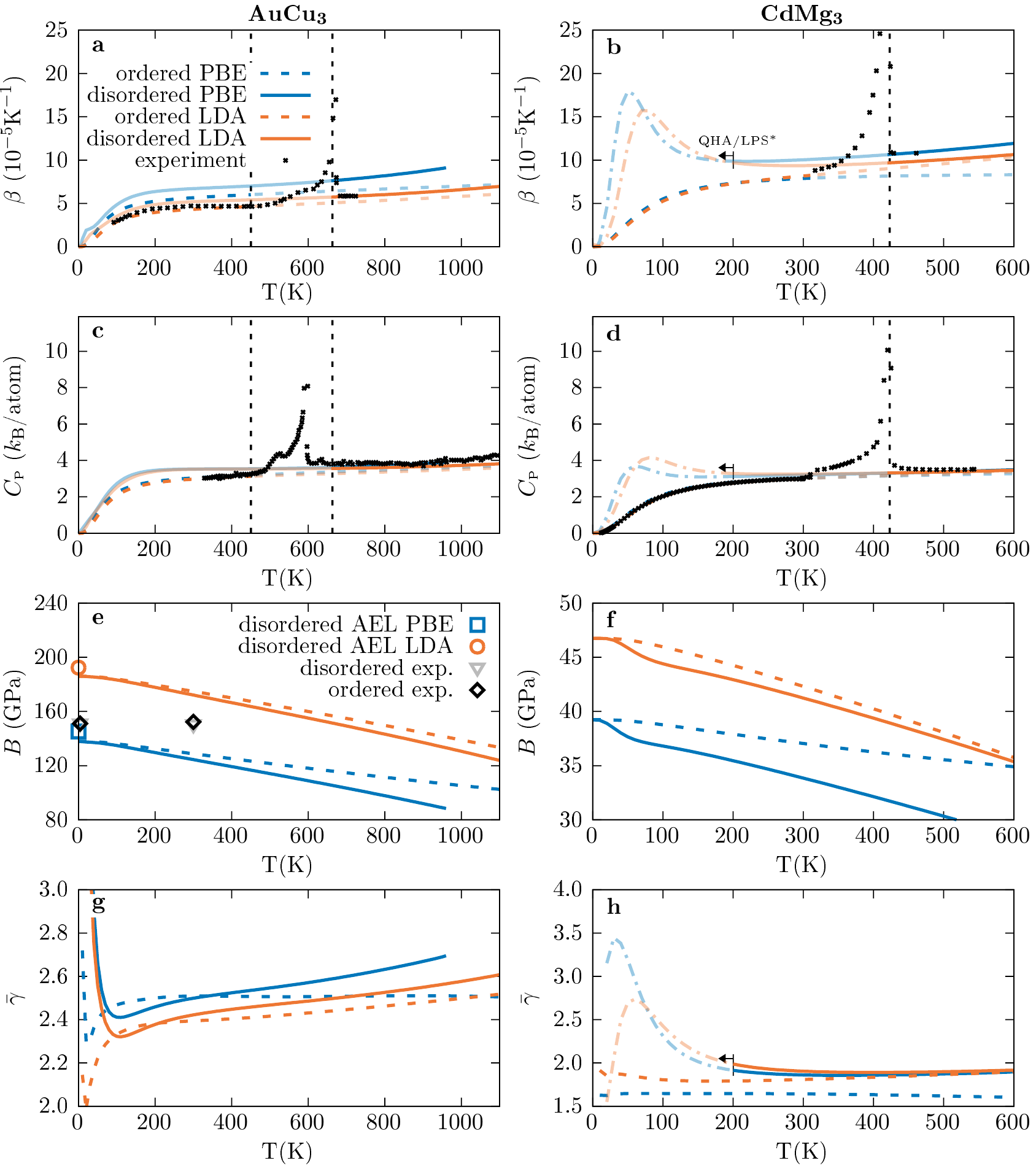}
  \vspace{-2mm}
  \caption{
    {\bf Thermomechanical properties of Cu$\mathbf{_3}$Au and CdMg$\mathbf{_3}$ from \QHPOCC.}
    {\bf a, b} Thermal expansion coefficient $\beta$, {\bf c, d} isobaric heat capacity $\CP$,
    {\bf e, f} bulk modulus $B$, and {\bf g, h} effective Gr\"{u}neisen parameter $\bar\gamma$
    of AuCu$_3$~(left) and CdMg$_3$~(right).
    The dashed vertical lines for AuCu$_3$ represent the transition temperatures to the partially
    and the fully disordered states, respectively. The dashed vertical lines for CdMg$_3$ denote
    the transition temperature to the disordered structure. Experimental values are taken from
    Refs.~\cite{Nix_PR_1941,Hirabayashi_JAP_1957,Hovi_ActaMetal_1964,Coffer_Wallace_JACS_1954,Johnston_Wallace_JACS_1957,Flinn_JPCS_1960}.
    \textcolor{black}{
      Note ``\QHA/\LPS''$^*$.
      For CdMg$_3$, the
      unphysical peaks of thermal expansion, heat
      capacity and Gr\"{u}neisen parameter, suggest the formation of long-period
      superstructures (\LPS), modulations of the hexagonal ordered structure
      D0$_{19}$ taken as reference for the \POCC\ expansion. The
      finite-size cutoff of the tiles can not represent the divergence
      of the characteristic lengths of the long-period
      superstructures. Thus, the constrained \LTVC\
      population vector \cite{aflowLTVC} will rotate and promote
      modifications of D0$_{19}$ containing antiphase boundaries \cite{Ghosal_ACA_2003} (with
      increasing volume) without reaching full \LPS\ description.}
  }
  \label{fig:fig2}
\end{figure*}

{\bf Calculations' details}.
\AFLOW\ \cite{aflowPAPER2023} leverages the Vienna \textit{ab initio} simulation package~(\VASP)
with all calculation parameters following the \AFLOW\ standard~\cite{curtarolo:art104}.
Exchange and correlation were treated with the projector augmented wave~(\PAW)
method~\cite{PAW} in either the local density approximation~(\LDA)~\cite{Ceperley_prl_1980} or
generalized gradient approximation~(\GGA) proposed by
Perdew, Burke, and Ernzerhof~(\PBE)~\cite{PBE}.
The cutoff energies are chosen to be 1.4 times the recommended maximum cutoffs~(\ENMAX)
of all pseudopotentials as set by \VASP.

Disordered compounds were modeled using the \AFLOWPOCC\ module
with a supercell size of 4, resulting in 7 and 29 unique {tiles}, respectively.
Phonon calculations were performed using the \AFLOWAPL\ module~\cite{curtarolo:art180}.
Supercell sizes and $\mathbf{k}$-point mesh dimensions were selected such that free energies were
converged to within 1~meV/atom for representative {tiles} of the disordered materials.
For ordered representatives of CdMg$_3$, the supercells used to calculate the phonon properties were
constructed to have at least 64 atoms, while for the ordered representatives of AuCu$_3$, the sizes
were chosen as specified in Table~\ref{tab:Cu3Au_parameters}.
Forces within these supercells were calculated with a $\mathbf{k}$-points mesh having at least
6,900 and 3,000 $\mathbf{k}$-points per reciprocal atom~(\KPPRA) for AuCu$_3$ and CdMg$_3$,
respectively~\cite{curtarolo:art104}.

\begin{table}[h]
  \caption{\small
    Parameters for the \QHA\ calculations of the ordered representatives of AuCu$_3$.
    The volume range is given in percent and in the format
    \textit{initial}:\textit{final}:\textit{increment}. Negative values denote compression, positive
    values expansion of the cell.
    \RHL: rhombohedral, \ORC: orthorhombic, \MCLC: base-centered monoclinic, \TET: tetragonal,
    \BCT: body-centered tetragonal, \ORCC: base-centered orthorhombic, \CUB: simple cubic.
  }
  \label{tab:Cu3Au_parameters}
  \begin{tabular*}{\columnwidth}{c @{\extracolsep{\fill}} cccc}
  \toprule
  \POCC\ structure & supercell & lattice type & {\color{blue}volume change (\%)} \\
  \colrule
  1 & $2 \times 2 \times 2$ & \RHL  & -12:12:6\\
  2 & $2 \times 2 \times 1$ & \ORC  & -12:12:6\\
  3 & $2 \times 2 \times 1$ & \MCLC &  -3:12:6\\
  4 & $3 \times 3 \times 2$ & \TET  & -12:12:6\\
  5 & $2 \times 2 \times 2$ & \BCT  & -12:12:6\\
  6 & $2 \times 2 \times 4$ & \ORCC &  -6:12:6\\
  7 & $2 \times 2 \times 2$ & \CUB  & -12:12:6\\
  \botrule
  \end{tabular*}
\end{table}

For the \EOS\ fit, vibrational properties were calculated at various compressed and expanded
volumes, typically spanning -12\% to 12\% of the equilibrium volume at a 6\% increment.
For AuCu$_3$, adjustments needed to be made to accommodate for the presence of imaginary
frequencies~(see Table~\ref{tab:Cu3Au_parameters}). Ordered representative \#3 of AuCu$_3$ has
unstable modes with the \LDA\ functional at 3\% volume compression, but these modes were easily
omitted in the calculation of the thermomechanical properties because their contributions to the
phonon \DOS\ were negligible. Ordered representative \#4 of AuCu$_3$ was found to be dynamically
unstable and was discarded entirely from the ensemble of structures. The calculated free
energies were fitted using the Stabilized Jellium \EOS~\cite{Teter_PRB_1995,Alchagirov_PRB_2001}:
\begin{equation}
  \label{eq:EOS_SJ}
  F(V) = \sum_{i = 0}^{3} f_i V^{-\frac{1}{3}i}.
\end{equation}
The parameters $f_i$ were determined using a polynomial fit.

\section*{Results and discussion}

\noindent
Two disordered alloys are used to validate the \QHPOCC\ method: AuCu$_3$ and CdMg$_3$.
These compounds were chosen due to their simple crystal structures and readily
available experimental data. AuCu$_3$ is ordered at low temperatures and crystallizes in the L1$_2$
structure~(\AFLOW\ prototype \href{https://aflow.org/prototype-encyclopedia/AB3_cP4_221_a_c.html}{AB3\_cP4\_221\_a\_c}~\cite{anrl1,anrl2,anrl3}).
Above 450~K, it becomes partially disordered and then fully disordered above 663~K, where it adopts
a face-centered cubic crystal structure~(\AFLOW\ prototype
\href{https://aflow.org/prototype-encyclopedia/A_cF4_225_a.html}{A\_cF4\_225\_a})~\cite{Nix_PR_1941,Keating_JAP_1951}.
CdMg$_3$ transitions from its ordered D0$_{19}$ phase~(\AFLOW\ prototype
\href{https://aflow.org/prototype-encyclopedia/A3B_hP8_194_h_c.html}{A3B\_hP8\_194\_h\_c})
to a hexagonal close-packed disordered structure at 423~K~\cite{Hovi_ActaMetal_1964,PhysRevB.48.748}.

\Figure~\ref{fig:fig2} shows the calculated thermomechanical properties for these materials in their
ordered and disordered phases using the Local Density Approximation~(\LDA) and the Perdew-Burke-Ernzerhof~(\PBE)
functional.
For both ordered and disordered AuCu$_3$, $\beta$ increases with
increasing temperature until they plateau above 200~K~(\Figure~\ref{fig:fig2}a).
The \CTE\ of the disordered structure is larger than of the ordered phase for all temperatures,
which is consistent with experiment. \PBE\ overestimates experimental $\beta$ values for both states.
The disordered phase even appears to diverge at higher temperatures, which is expected for \QHA,
as anharmonic contributions become increasingly dominant. \LDA\ shows excellent agreement with
experimental values, suggesting that the choice of the functional is the reason for the bad
agreement for \PBE.

The \CTE\ of CdMg$_3$ is shown in Figure~\ref{fig:fig2}b. The curves of the ordered phases show
similar behavior as those for AuCu$_3$, except for $\beta$ for \LDA, which converges towards the
value of the disordered phase.
Good experimental values for the ordered phase are unavailable in literature, but extrapolating
from the tail found in the experimental values, it is anticipated to see good agreement.
Above the experimental order-disorder transition temperature, the disordered phase shows good
agreement for \PBE\ whereas \LDA\ underestimates $\beta$. This shows that the functional choice is
critical when calculating thermal expansion of these materials.
\QHPOCC\ also shows a striking difference in the disordered phase compared to AuCu$_3$: $\beta$ of
disordered CdMg$_3$ exhibits a peak at 50\,K and 70\,K for \PBE\ and \LDA, respectively.

The isobaric heat capacities shown in \Figure~\ref{fig:fig2}c for AuCu$_3$ and in
\Figure~\ref{fig:fig2}d for CdMg$_3$ are less sensitive to the chosen functional. The agreement with
experimental values for the ordered structures are excellent for both functionals. $\CP$ for the
disordered phases are underestimated, but still in good agreement with experiments. A peak can be
observed in \QHPOCC\ for both \LDA\ and \PBE\ in disordered CdMg$_3$, with the peak for \LDA\ being
at a higher temperature.

For the bulk modulus, experimental data is only available for AuCu$_3$~\cite{Flinn_JPCS_1960}. As
\Figure~\ref{fig:fig2}e shows, \PBE\ underestimates $B$ whereas \LDA\ strongly overestimates it.
$B$ is very sensitive to the chosen functional since it depends on the volume of the unit cell. This
is not a shortcoming of \QHPOCC, but of \QHA, as this behavior was also found for simple ordered
materials such as MgO and CaO~\cite{Erba_Dovsei_JCP_2015}. To further substantiate that \QHPOCC\ is
consistent with other methods, $B$ was calculated at 0\,K using the \AFLOW\ Automatic Elastic
Library~(\AEL)~\cite{curtarolo:art115} --- \Figure~\ref{fig:fig2}e demonstrates that \AEL\ and
\QHPOCC\ are consistent with each other. The values for the disordered phase is smaller than for the
ordered phase at all temperatures, which is consistent with experiments. However, $B$ drops much
faster with temperature than experiments suggest, which has also been observed in ordered
materials~\cite{Erba_Dovsei_JCP_2015}.

As for AuCu$_3$, CdMg$_3$ has a smaller bulk modulus in the disordered than in the ordered state,
with \LDA\ resulting in larger moduli than \PBE. The temperature dependence of disordered CdMg$_3$
differs from AuCu$_3$ by showing an inflection, the end point of which coincides with the
temperature of the peak in $\beta$ and $\CP$ for both functionals.

Finally, \Figures~\ref{fig:fig2}g and h show the effective Gr\"{u}neisen parameters for both
materials. In general, $\bar\gamma$ is larger for the disordered material than the ordered material,
particularly at higher temperatures. An exception is \LDA\ and CdMg$_3$, where $\bar\gamma$ for the
ordered phase approaches the disordered phase due to the divergence of $\beta$ at higher
temperatures.
Higher values of $\bar\gamma$ correlate with lower thermal conductivity, which is
expected for materials with substitutional disorder~\cite{Braun_ESO_AdvMat_2018}. The divergences
found at low temperatures are the result of the calculation method: as Equation~\ref{eq:Grueneisen}
shows, $\bar\gamma$ depends on the quotient of $\beta$ and $\CV$, both of which are very small at
low temperatures, causing numerical instabilities. This relationship also causes the peak for
CdMg$_3$ as $\beta$ peaks stronger than $\CP$.

In experiments, peaks in thermal expansion and isobaric heat capacity typically indicate phase
transitions. Here, peaks are not found in \QHA\ for ordered phases and in
disordered CdMg$_3$ regions.
\textcolor{black}{
  From the solid solution and upon reducing temperature, CdMg$_3$ transitions from a disordered hexagonal close-packed
  structure to the ordered D0$_{19}$ \cite{Massalski,PhysRevB.48.748}. With further temperature
  reduction, D0$_{19}$ often develops antiphase boundaries and complex stacking faults \cite{Ghosal_ACA_2003},
  associated with forming long-period superstructures,
  \LPS.
  Notably, several Cd and Mg binary alloys show similar features \cite{Massalski,monster,curtarolo:art54}.
  The divergence of the characteristic length scale in the \LPS\ can not
  be reproduced by the tiles having a finite size, making
  the \POCC\ expansion incomplete. Therefore, the constrained \LTVC\
  population vector \cite{aflowLTVC} will rotate and promote
  modifications of D0$_{19}$ containing antiphase boundaries and
  stacking faults having increased volume without ever reaching full
  \LPS\ description.
  While such computational limitation is expected to
  appear at low temperatures and in all systems having ordered ground-states with
  characteristic length scales too large for the \POCC-tiles expansion,
  our algorithm has been devised to study disordered systems at high
  temperatures where only short-range needs to be described \cite{curtarolo:art80,curtarolo:art187}. As such, the
  limitation does not pose any concern.}

\begin{figure}
  \includegraphics{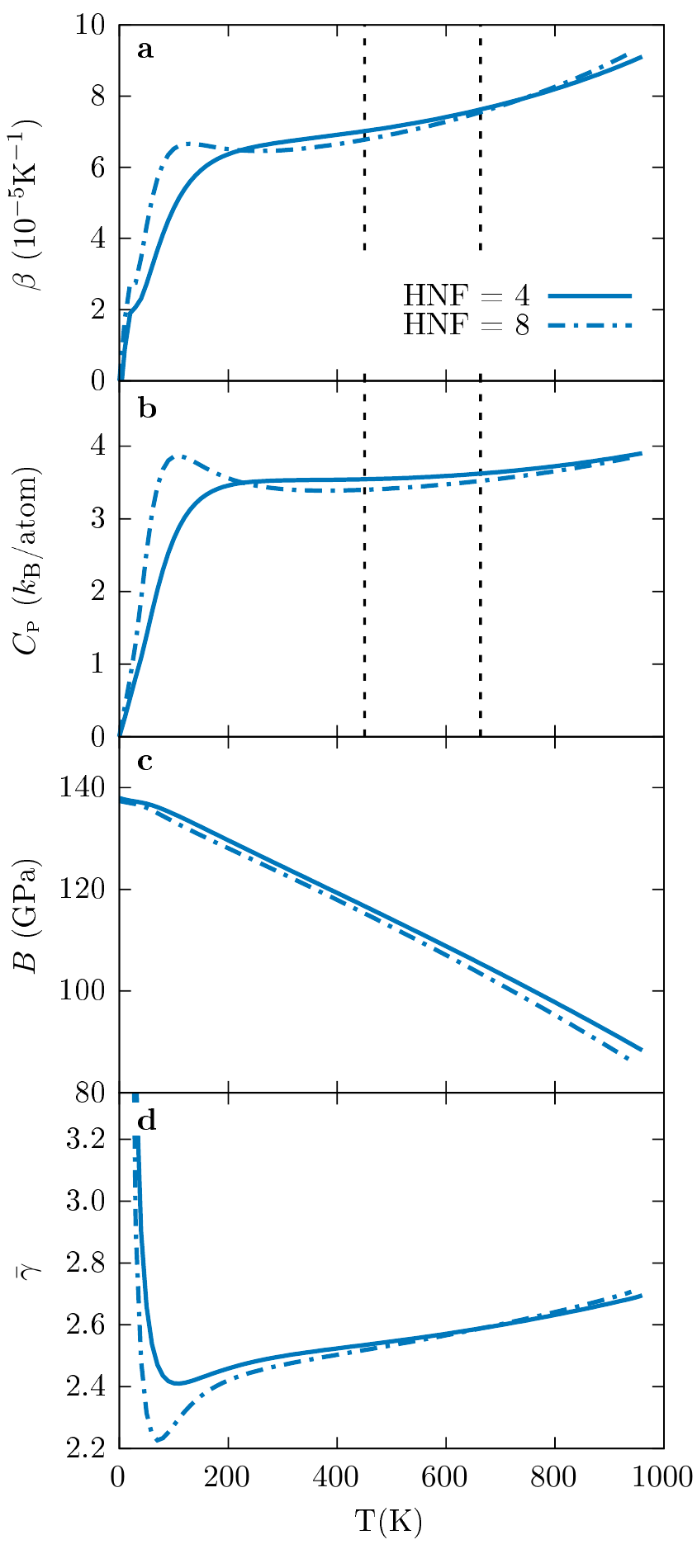}
  \caption{
    {\bf Thermomechanical properties of Cu$\mathbf{_3}$Au for larger \POCC\ supercells.}
    {\bf a} Thermal expansion coefficient $\beta$, {\bf b} isobaric heat capacity $\CP$, {\bf c}
    bulk modulus $B$, and {\bf d} effective Gr\"{u}neisen parameter $\bar\gamma$ of AuCu$_3$ for
    \POCC\ supercell sizes~(\HNF) 4~(solid lines) and 8~(dash-dotted lines).
    The dashed vertical lines represent the transition temperatures to the partially and the
    fully disordered state, respectively.
  }
  \label{fig:fig3}
\end{figure}

Experiments suggest that a peak should appear in $\beta$ and $\CP$ for AuCu$_3$ as well, but none
are observable in \Figures~\ref{fig:fig2}a and c. The reason for this discrepancy could be the
sample size of the canonical ensemble: \POCC\ creates only 7 ordered representatives for AuCu$_3$,
but 29 for CdMg$_3$. To test this hypothesis, the size of the \HNF\ matrix was increased to 8,
which produced 49~ordered representatives.
As \Figures~\ref{fig:fig3}a and b show, increasing the maximum \POCC\ tile size
results in a peak in both thermal expansion and heat capacity, confirming that small sampling
suppressed the appearance of the phase transition.
However, no inflection point can be observed in the bulk modulus as was observed in
CdMg$_3$~(see \Figure~\ref{fig:fig3}c). This is not surprising because the values for the ordered
and disordered phase are similar at low temperatures in both \QHPOCC\ and experiments, making any
inflection imperceptible. Finally, the effective Gr\"{u}neisen parameter, shown in
\Figure~\ref{fig:fig3}d, shows a deep dip at lower temperatures when the supercell size is increased
due to the larger peak height in $\CP$, which is in the denominator of $\bar\gamma$.

As with CdMg$_3$, the calculated transition temperature of AuCu$_3$ is far below the experimental
value. While transitioning from an ordered to a disordered phase, both materials are in a partially
ordered state spanning at least 100~K.
The requirement of \POCC\ that the stoichiometry of the tiles correspond to the disordered structure
parametrizes this transition region as a line.
To improve the accuracy of the transition temperature, a more complete picture of
the tile distribution is needed so that the variation of $S_{\rm tiling}$ can be captured.
This could be achieved, for example, by including structures with compositions different from
the parent structure.
A similar approach is employed in the Lederer-Toher-Vecchio-Curtarolo~(\LTVC) method, which can
accurately predict the transition temperature, but requires more input data~\cite{aflowLTVC}.

Outside the transition region, however, all properties are well converged with respect to the \POCC\
supercell size.
So, while $\Omega_{\rm tiling}$ and $S_{\rm tiling}$ may change with increased
maximum tile size, their temperature derivatives do not.
The smallest supercell is thus sufficient to calculate thermophysical properties of the
disordered material.

\section*{Conclusions}
\noindent
This work introduces a new framework, \QHPOCC, to perform finite temperature calculations of
thermophysical properties of systems with substitutional disorder.
The method uses the Partial OCCupation algorithm and the quasi-harmonic approximation~(\QHA) to
calculate finite-temperature free energies using a canonical ensemble. This ensemble is then used to
determine properties accessible via the \QHA.
Two systems, AuCu$_3$ and CdMg$_3$, were used to validate the approach by comparing calculated
values of thermal expansion and isobaric heat capacity with experimental data.
The results are in a good agreement with experiments, demonstrating that \QHPOCC\ is a promising
method to study finite-temperature vibrational properties.
Due to the implicit incorporation of the configurational energy, \QHPOCC\ is able to recover
features that are inaccessible to methods that rely on only one supercell. Peaks in both thermal
expansion and heat capacity indicate structural phase transitions.
The combination of computational efficiency,
capability to bypass the calculation of the tiling-entropy,
and overall accuracy makes \QHPOCC\ an
excellent screening tool for thermomechanical properties of disordered materials in the
quasi-harmonic approximation.

\section*{Declaration of Competing Interest}
\noindent
The authors declare no competing interests.

\section*{Software and Data Availability}
\noindent
{\bf Code.}
The code is freely available as part of the \AFLOW\ software suite \cite{aflowPAPER2023}. All information to reproduce
the data are documented in this article.\\
{\bf Data.}
All structure data is freely available and accessible online through
AFLOW.org \cite{aflowlibPAPER2023} or programmatically via the REST- \cite{aflowAPI} and AFLUX Search-APIs \cite{aflux}.
The AFLOW prototype information is provided online at
\href{https://aflow.org/prototype-encyclopedia}{http://aflow.org/prototype-encyclopedia}~\cite{anrl1,anrl2,anrl3},
and the corresponding structures can be generated with the AFLOW
source code

\newcommand{\Ozolins}{Ozoli{\c{n}}{\v{s}}}

\end{document}